\documentclass{article}

     \PassOptionsToPackage{square,sort,comma,numbers}{natbib}

     \usepackage[preprint]{neurips_2019}

\usepackage[utf8]{inputenc} 
\usepackage[T1]{fontenc}    
\usepackage{booktabs}       
\usepackage{amsfonts}       
\usepackage{nicefrac}       
\usepackage{microtype}      
\usepackage{subcaption}
\usepackage{ifthen}
\usepackage{color}
\usepackage{glossaries}

\usepackage{graphicx}
\usepackage{multirow}
\usepackage{xspace}
\usepackage[dvipsnames]{xcolor}
\usepackage{enumitem}
\usepackage{wrapfig}

\usepackage{amsmath}
\usepackage{amssymb}
\usepackage{mathtools}

\DeclareMathOperator{\argmax}{argmax}
\DeclareMathOperator{\argmin}{argmin}

\usepackage{array}
\newcolumntype{L}[1]{>{\raggedright\let\newline\\\arraybackslash\hspace{0pt}}m{#1}}
\newcolumntype{C}[1]{>{\centering\let\newline\\\arraybackslash\hspace{0pt}}m{#1}}
\newcolumntype{R}[1]{>{\raggedleft\let\newline\\\arraybackslash\hspace{0pt}}m{#1}}
\newcolumntype{P}[1]{>{\raggedright}p{#1}}

\definecolor{urlBlue}{HTML}{2196F3}

\title{Human-Machine Collaboration \\ for Fast Land Cover Mapping}

\author{
    Caleb Robinson$^{1,}$\thanks{Work done while interning at Microsoft Research} \quad Anthony Ortiz$^{2,}$\footnotemark[1] \quad Kolya Malkin$^{3}$ \quad Blake Elias$^{4}$ \quad Andi Peng$^{4}$ \\ {\bf Dan Morris$^{4}$ \quad Bistra Dilkina$^{5}$ \quad Nebojsa Jojic$^{4,}$\thanks{\texttt{jojic@microsoft.com}}} \\
    $^1$Georgia Institute of Technology \qquad $^2$University of Texas at El Paso \qquad $^3$Yale University \\ $^4$Microsoft Research \qquad $^5$University of Southern California
}

\begin{document}

\maketitle

\begin{abstract}
We propose incorporating human labelers in a model fine-tuning system that provides immediate user feedback. In our framework, human labelers can interactively query model predictions on unlabeled data, choose which data to label, and see the resulting effect on the model's predictions. This bi-directional feedback loop allows humans to learn how the model responds to new data. Our hypothesis is that this rich feedback allows human labelers to create mental models that enable them to better choose which biases to introduce to the model. We compare human-selected points to points selected using standard active learning methods. We further investigate how the fine-tuning methodology impacts the human labelers' performance. We implement this framework for fine-tuning high-resolution land cover segmentation models. Specifically, we fine-tune a deep neural network -- trained to segment high-resolution aerial imagery into different land cover classes in Maryland, USA -- to a new spatial area in New York, USA. The tight loop turns the algorithm and the human operator into a hybrid system that can produce land cover maps of a large area much more efficiently than the traditional workflows.  Our framework has applications in geospatial machine learning settings where there is a practically limitless supply of unlabeled data, of which only a small fraction can feasibly be labeled through human efforts.
\end{abstract}

\section{Introduction} \label{sec:introduction}
Machine learning models are usually imagined as artificially ``intelligent'' agents that mimic human autonomy and generalization abilities: having explored their training environment, they are supposed to choose their actions independently and reliably in similar situations. While this notion of intelligence guides the design and testing of new algorithmic ideas, in practice, the resulting algorithms are rarely treated as capable of either autonomy or generalization. Instead, human hand-holding is present throughout a AI model's development and deployment: researchers and engineers acquire data with a specific goal in mind, then work on finding and tuning the methods that handle the peculiarities of the data well. When the algorithm is eventually deployed, it often suffers from domain shift, where slight changes in the statistics of real-world input compared to the training input can degrade performance considerably. Thus, the algorithm is constantly reevaluated through human monitoring, which may trigger a process requiring repeated data acquisition and retraining. Hence, most practical deployments are better thought of as examples of hybrid -- rather than purely artificial -- intelligence. Active learning loops can be seen as an approximate model  of such hybrid human-machine intelligence, as long as humans are allowed deeper involvement than just as labeling oracles.
We focus on one such hybrid application: a process for land cover mapping. We investigate whether it is possible to maximize performance in the land cover mapping task by directly integrating humans into the training loop instead of isolating the artificially intelligent component.

\begin{figure*}[t]
    \centering
    \includegraphics[width=0.8\textwidth]{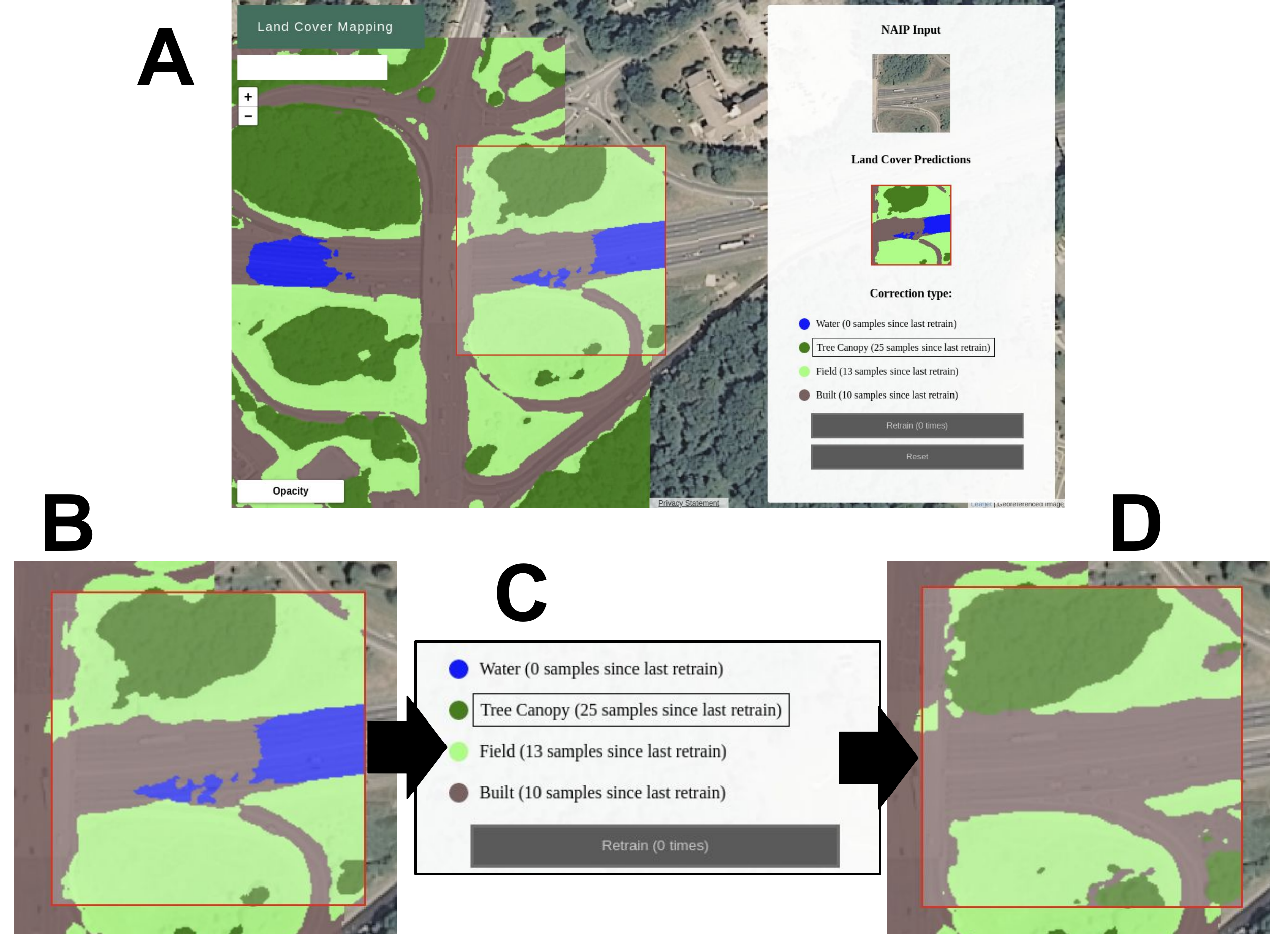}
    \caption{User interface of our land cover labeling tool. (See the video - download link can be found above Appendix A) (\textbf{A}) Land cover prediction results are overlayed on top of the map. (\textbf{B}) The user can easily identify misclassified pixels and (\textbf{C}) submit corrections by clicking on the map. (\textbf{D}) Pressing ``Retrain'' updates the model and displays new land cover predictions in the interface. In this example, the user provided a handful of point corrections in the impervious surface initially misclassifed as water.}
    \label{fig:web-tool}
\end{figure*}

We summarize our main contributions as follows:
\begin{itemize}
    \item We design an \textbf{interactive web tool} that allows users to test a high-resolution land cover model on any patch of land on a satellite map, then -- in the same interface -- relabel pixels of their choosing and retrain (``fine-tune'') the model (see Fig.~\ref{fig:web-tool}). 
    \item In an online user study, we show that users develop a \textbf{theory of mind} of the learning system that improves the performance of the hybrid human-AI intelligence over time, and that the hybrid intelligence increases users' \textbf{trust in the AI}. 
    \item We study the effectiveness of the combination of different active learning \textit{query methods} with different model \textit{fine-tuning methods} in an offline study and find that querying for labels at \textbf{randomly} selected points outperform or nearly matches standard active learning \textit{query methods} (see Fig.~\ref{fig:offline-results}).
    \item We examine how well \textbf{human labelers} function as sample \textit{query methods} as compared to automatic selection methods. We find that humans perform significantly better, even compared to learning systems in which the model is told on which points it is making labeling errors (see Fig.~\ref{fig:online-results}). Furthermore, the value of human-provided labels increases with the time humans spend using the tool.
\end{itemize}


\section{Background}

A traditional active learning setup consists of a parametrized model, an unlabeled `pool' of data, a data  \textit{query method} (also known as the \textit{data selection method}), and a \textit{labeling oracle}. One iteration of fine-tuning the model consists of utilizing the query method to choose data points for labeling, a labeling oracle to provide those labels, and fine-tuning the model parameters to these additional data samples.

Here, humans' involvement is deliberately minimized in order to create a more autonomous artificially intelligent agent: Humans only act as labeling oracles. In practice, a deployed system may suffer from a host of issues~\cite{settles2011theories} that require continual effort and redesign from the humans that are hidden ``outside the loop''.\footnote{\cite{settles2011theories} lists six practical challenges for active learning systems including: noisy oracles, variable labeling costs, and changing (or unknown) model classes.} In our work, we inject humans into the learning loop in both the selection \emph{and} labeling of new training samples for model adaptation, functioning as both the data \textit{query method} and \textit{labeling oracle}.

The purpose of the \textit{query method} is to pick unlabeled data that, when labeled, will provide the largest benefit to the model. In active learning, the learner is allowed to ask for help by querying the label oracle, but it must know \emph{when} to ask for help. Conventional approaches \cite{zhang2017active,settles2012active} ask the oracle to label instances with low prediction confidence. Another recent approach models uncertainty in labeling oracles \cite{huang2016active}. 
Unfortunately, the algorithms often suffer from ``unknown unknowns'': self-inspection does not reliably reveal what is not modeled well. This is the case in most ML algorithms, including deep neural networks~\cite{nalisnick2018do}. 
On the other hand, by observing the effects of their decisions on a model being retrained on-the-fly, human labelers are expected to adapt their own data selection process to reflect not only their understanding of the data, but also their developing intuition regarding the inner workings of the model and its adaptation algorithms.

\begin{figure}[ht]
    \centering
    \includegraphics[width=0.9\textwidth,trim=0 0 0 15]{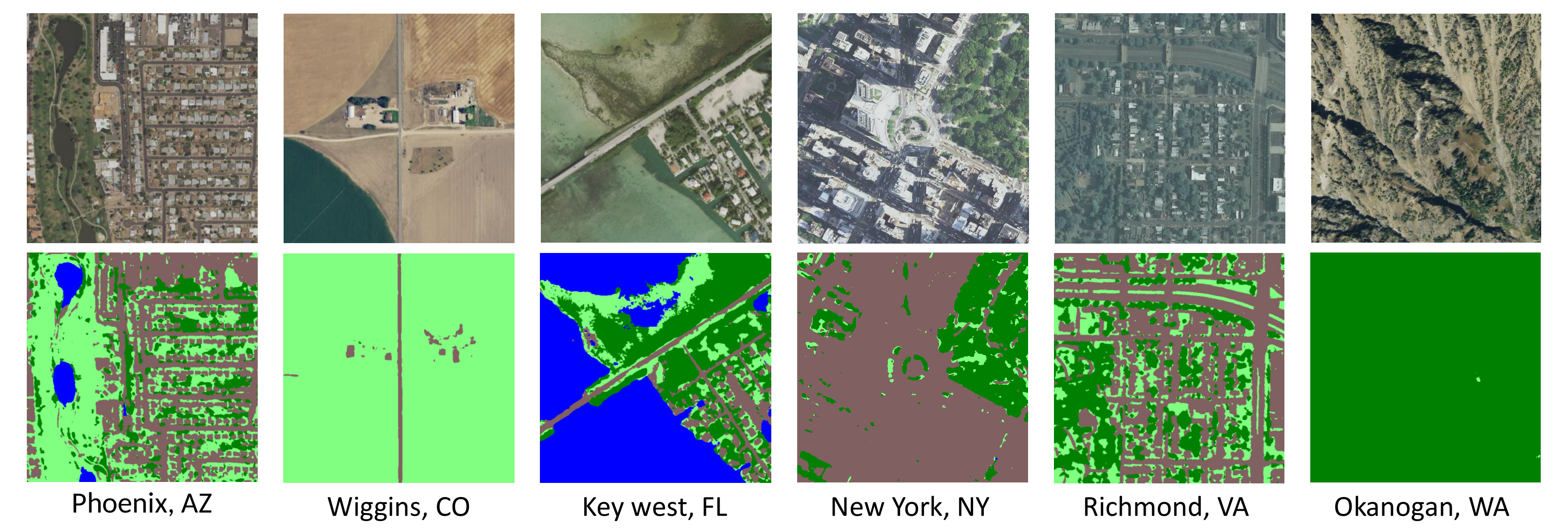}
    \caption{National Agriculture Imagery Program (NAIP) aerial imagery (\textbf{top row}) with modeled land cover estimates (\textbf{bottom row}). Existing supervised learning models, trained for generating land cover labels from aerial imagery, do not generalize well due to the large spatial and temporal variances in aerial imagery. Creating accurate land cover maps at a massive scale therefore requires additional human interventions. We propose an interactive model fine-tuning system, coupling human labelers and machine learning models, for facilitating these interventions.}
    \label{fig:teaser}
\end{figure}

Land cover mapping -- the segmentation of aerial or satellite imagery into land cover classes such as ``water'', ``forest'', ``field/low vegetation'', or ``impervious surface'' (Fig.~\ref{fig:teaser}) -- has attracted reinvigorated interest in machine learning research \cite{pal2005random,robinson2019large, demir2018deepglobe,rakhlin2018land,tian2018dense,kuo2018deep}. High-resolution land cover maps are an essential component in environmental science, agriculture, forestry~\cite{hansen2013high}, urban development~\cite{zhang2013analysis}, the insurance and banking industries, and for demography in developing countries~\cite{fb2019pop}. Satellite imagery is being produced on an increasingly frequent basis. However, despite their importance, high-resolution land cover maps are not yet widely available as neither ML algorithms nor human labor scale appropriately \cite{robinson2019large}.

{\bf To a machine learning or computer vision researcher}, land cover mapping is a semantic segmentation problem. 
Machine learning models are not yet able to generate high-resolution (1m / pixel) land cover labels with an accuracy that matches human labeling. A major obstacle is that high-resolution land cover labels for training such models only exist in small, specialized locations \cite{demir2018deepglobe,rakhlin2018land,dstldataset,yang2010bag, castelluccio2015land,ISPRS}. In \cite{robinson2019large}, it is shown that a state-of-the-art deep neural network trained on 1m-resolution images and labels from a much larger (160,000~km$^2$) dataset~\cite{chesapeakeData1, chesapeakeData2} in the Chesapeake Bay watershed (north-eastern US) still does not perform well in the mid-western US. Other recent work also utilizes additional, more broadly available input data \cite{kampffmeyer2018urban,malkin2018label}. However, all existing land cover models are biased by the geographic locations on which they were trained. Large systematic errors in predictions limit their applicability and are challenging to detect at scale.\footnote{For example, imagery of the contiguous US at 1m resolution covers 8 trillion pixels.}
Finally, the classification tasks are constantly shifting. While one dataset may segment vegetation simply into ``low vegetation'' and ``tree canopy'', other applications may require delineating coffee farms from orchards.

Therefore, land cover mapping offers an opportunity to apply domain transfer/adaptation, few-shot learning, and meta-learning research~\cite{yosinski2014transferable, tuia2009active,finn2017model,snell2017prototypical, vinyals2016matching,ravi2016optimization}. In most of this research, however, ground truth labels are available in target domains, and the models are evaluated based on how sample-efficient they are in left-out datasets \cite{finn2017model,snell2017prototypical, lake2015human}. In GIS applications, generic algorithms are defective at high-resolutions, and the best-performing algorithms pre-trained on one geographic area still require additional data in a new area to achieve optimal performance~\cite{malkin2018label, robinson2019large}. 

{\bf To a Geographic Information Systems (GIS) professional}, however, land cover mapping is an inherently human-driven process augmented by technology. Accurate and useful labels themselves, not a training dataset for ML algorithms, are the immediate goal.  The process typically starts with color-based segmentation algorithms that create initial maps, followed by experts who provide labels in different areas, creating rules on the fly, and then manually correcting the remaining errors. The labor efficiency of the process may increase as the humans learn how to use these tools better, but is not boosted by quick adaptation of the classification algorithms themselves.\footnote{Typically, separately tuned random forests are used, although neural networks are rapidly gaining traction.} This makes land cover mapping at the resolution and scale needed today cost-prohibitive for most agencies.

{\bf A hybrid system for accurate and efficient land cover labeling} would more tightly integrate the human and machine efforts. Here we investigate a land cover mapping workflow where users' work immediately affects the performance of prediction algorithms.
Our design, which incorporates human feedback integrated in real time as training points for our model, can be seen as an instance of machine teaching \cite{simard2017machine, zhu2015machine}, as humans deploy their own intelligence to identify and correct mislabeled points in an effort to improve the model. However, our system does not attempt to create an autonomous entity, capable of generalizing, as the final result: the ability to efficiently label large areas is the goal, and the final trained algorithm is but one aspect of the overall workflow.
To a human, the ML model is simply a powerful macro that they (re)define on the fly in order to amplify their work. To the ML model, the human is the source of data to learn from. Together, this hybrid system holds the potential to outperform existing GIS workflows as well as pure ML approaches in cost and accuracy.

\section{Land cover study design}
\label{sec:study_design}

We focus on the following task to be performed by a combination of algorithmic and human inference: given a pretrained segmentation network, which was trained on a 1m-resolution, four-class land cover map of Maryland~\cite{chesapeakeData1}, we would like to quickly (within at most 15 minutes) produce accurate maps of  12km $\times$ 7km regions of 1m-resolution imagery in New York State. The study targets four regions at various levels of urban development. We aim to create the map of each region by slightly changing the parameters of the Maryland model to fit a limited number of guidance points in the new areas.%

We vary two parameters in our study: the \textbf{fine-tuning method} and \textbf{query method}.

The \textbf{fine-tuning method} is the algorithm for retraining the model to fit new guidance points. Such a method needs to be fast and sample-efficient. As we have ground truth data in the entire Chesapeake watershed, including Maryland and N.Y., these choices can be studied offline. Extensive comparisons of various methods, described in Sec.~\ref{sec:experiments_offline}, led us to choose two methods to test in online user studies: 
\begin{description}
\item [\bf Last 1 Layer] The full adaptation of the 64 $\times$ 4 parameters in the last (softmax) layer (gradient descent to convergence on all user-supplied points).
\item [\bf Last 2 Layers] A fixed number of iterations of gradient descent on the parameters of the last two layers.
\end{description}

The \textbf{query method} is the method for selecting guidance points on which to fine-tune the model on a new region. The main object of our study is to compare automatic methods, such as random selection or active learning approaches, to \textbf{hybrid (human-guided) methods}, where users iteratively view the current model's predictions, correct the labels at points of their choice, and trigger model retraining. The traditional active learning approaches to automatic selection of points to query can also be studied offline on a fully labeled dataset (Sec.~\ref{sec:experiments_offline}).

To study the human-guided approach, we developed a web tool for including humans in the process of land cover map creation. The tool allows users to iterate between labeling and testing the model (Fig.~\ref{fig:web-tool}). The tool exploits the spatial nature of the data in the task, allowing the user to zoom and pan in the high-resolution imagery to find areas where they want to test the current algorithm. Upon a click on the map, the prediction of the current model on a surrounding 200m$\times$200m patch of land is overlaid on the map. The user can then label pixels of their choice, either where they see errors or for some other reason they think that the label will be useful. They can induce near-instant retraining of the model at any time with the click of a button. After that, they can check how well the retrained model works by clicking on the imagery again. 

In our study, we allow each user to spend 15 minutes in each of the four target areas. The order of the areas is randomized, and the model is reset to the pretrained baseline with each new area. Users are given both of the fine-tuning methods to work with, but are not told of this: each user is assigned to work with the same (randomly chosen) fine-tuning method in the first three tasks, and then the other method in the last task. Such an assignment of areas and methods allow us to separate the first task -- during which the user is getting used to the tool -- from tasks 2 and 3, where the user is assumed to be doing their best work, and from task 4, where the learning system changes its behavior. This allows us not only to measure the variation in performance across users, methods, and regions, but also to see if the users are building an understanding of how the model and its adaptation work.    

In addition to the main tool, we use a traditional crowdsourcing platform to obtain accurate ground truth labels for evaluating the adapted models in the new areas, as the semi-manually-created ground truth Chesapeake is only $90-95\%$ accurate. The user is shown a random patch of imagery within the four target areas and asked to label the central pixel (see Appendix B.1 for details). This tool does not provide the user with any communication with a trained model, isolating the human work from the machine work. The data obtained from this tool allows us to evaluate the performance of models created by users though the main tool and compare them to models trained using other query methods.

\section{Base model and training data}
We train models that take as input patches of high-resolution (1m) four-band aerial imagery from the USDA National Agriculture Imagery Program (NAIP) and predict high-resolution segmentations into four land cover classes (water, forest, field, impervious surfaces). The default training label datasets are from \cite{malkin2018label, chesapeakeData1}.

Our base model is a neural network that produces probabilities of each class at each image pixel. Given the model parameters $\theta$ and an image $X=\{x_{ijk}\}\in{\mathbb R}^{w\times h\times c}$ (where $c=4$ is the channel depth and $w\times h$ are the image dimensions), the model outputs a probability distribution over the target classes at each pixel, i.e., 
$f(\theta,X)\in{{\cal D}(n)}^{w\times h}$, where ${\cal D}(n)$ is the probability simplex on the $n=4$ output classes. This yields distributions over labels $P_\theta(\hat{y}_{ij}|X)$ for each coordinate $(i,j)$.

The network is similar to the U-net architecture \cite{ronneberger2015u,rakhlin2018land}. We trained the network on randomly selected patches sampled from the state of Maryland. The training settings match those of \cite{malkin2018label}; see Appendix A for details.

\section{Offline active learning  experiments} \label{sec:experiments_offline}

As discussed in Sec.~\ref{sec:study_design}, we investigated different methods for adapting a pre-trained model with newly acquired label data in different domains, working with four target areas in in New York, USA. Our offline experiments are meant to identify the optimal fine-tuning and query methods, which are then used in online user studies. In our offline experiments, the base model is adapted to a small number -- 10 to 2000 -- of \emph{automatically chosen} labeled pixels (less than 0.01\% of each target area). Then the performance is evaluated on the entirety of the target areas. The following fine-tuning methods were tested:

\begin{description}
\item[\textbf{Last $k$ Layers}] Following \cite{yosinski2014transferable}, the final $k$ convolutional layers in the U-net architecture have their weights exposed as trainable via gradient descent (initialized from the weights of the base model), while all other parameters in the network are held fixed. Here, $k \in \{1, 2, 3\}$.

\item[\textbf{Group normalization parameters}]  Inspired by the success of feature-wise transformations~\cite{dumoulin2018feature-wise} in neural style transfer~\cite{dumoulin2016learned} and visual question answering~\cite{perez2018film} we extended it for model fine-tuning.  Our U-net architecture uses group normalization \cite{wu2018group} in the final convolutional layers. The group normalization parameters affect large groups of filters in each layer via single affine transformation, with the assumption that filters within a group are correlated. Thus, training these parameters to fit new training points causes correlated changes in the layers' outputs, providing a regularized mechanism to affect the entire network, in contrast with full backpropagation, which affects all weights in the chosen layers.

\item[\textbf{Dropout}] We effect dropout, i.e., set the outputs of a fixed subset of the neurons to 0, in the final $k$ convolutional layers. Searching for the binary mask that minimizes a loss is a discrete optimization problem, which we solve using an elementary genetic algorithm. Here we use $k=5$ and a mean dropout rate of $0.2$, but we conducted only limited experiments due to the high cost of this method, which requires evaluation of the model at all sample points at each of 64 mutation iterations. However, we hypothesize that this highly constrained method is less prone to overfitting than techniques based on backpropagation.
\end{description}

Motivated by \cite{zhang2017active,settles2012active}, we also investigated three query methods for selecting the additional 10 to 2000 labeled pixels used by the fine-tuning methods:

\begin{description}

\item[\textbf{Random}] Sample points $(i^*,j^*)$ uniformly randomly from the training area.

\item[\textbf{Entropy}]  Select points which maximize the Shannon entropy of output distributions over classes:
\[(i^*,j^*) = \argmax_{(i,j)} \left(- \sum_\ell P_{\theta}(\hat{y}_{ij}=\ell|X) \log{P_{\theta}(\hat{y}_{ij}=\ell|X)}\right).\]

\item[\textbf{Min-Margin}] Select points which minimize the difference between probabilities assigned to the most-likely and second-most-likely classes: \[(i^*,j^*) = \argmin_{(i,j)} \left(P_{\theta}(\hat{y}_{ij}=\ell^1_{ij}|X) - P_{\theta}(\hat{y}_{ij}=\ell^2_{ij}|X)\right),\] where $\ell^1_{ij}$ and $\ell^2_{ij}$ are the two most likely classes under $P_\theta(\hat{y}_{ij}|X)$.
\end{description}

We also include the following method, the purpose of which is to make a comparison with \emph{humans} selecting mistake points in our online study. It is \textbf{not} an automatic query strategy, as it assumes the model has access to an all-knowing labeling oracle \emph{before} it chooses where to query the oracle for labels. It simply mocks a teacher that feeds randomly chosen mistake points to the model.
\begin{description}
\item[\textbf{Mistakes}] Uniformly sample points $(i^*,j^*)$ where the model's prediction disagrees with the ground truth.
\end{description}

Because it is prohibitively costly to select points using the \textbf{Entropy}, \textbf{Min-Margin}, and \textbf{Mistakes} methods at \emph{every} training iteration, we approximate this procedure by batching: periodically evaluating the model on the training area and selecting the optimal points among a large set of 10000 uniformly sampled positions. Namely, we evaluate the model and select a new batch of points after 10, 40, 100, 200, 400, 1000, and 2000 points have been chosen.

\begin{figure*}[t]
    \centering
    \begin{tabular}{cc}
    \includegraphics[width=0.49\textwidth,trim=0 0 0 0]{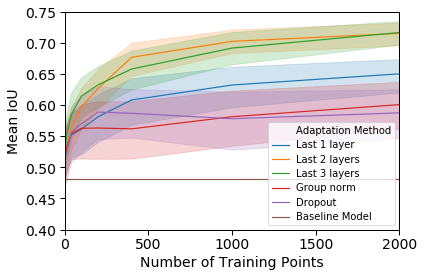} & 
    \includegraphics[width=0.49\textwidth,trim=0 0 0 0]{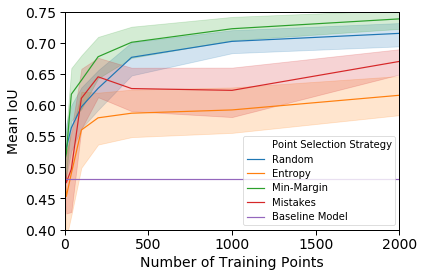} \\
    different fine-tuning methods & \textbf{Last 2 Layers} fine-tuning method\\
    \textbf{Random} query method & different query methods
    \end{tabular}
    \caption{Performance of different fine-tuning methods (left) and query methods (right), mean and standard deviation over 5 runs and 4 target areas. At several stages -- after 10, 40, 100, 200, 400, 1000, and 2000 points have been seen -- the system selects a further set of training points using the given query method and retrains the model using the fine-tuning method. The performance of the model evaluated on the entire target region tends to improve as more points are seen.}
    \label{fig:offline-results}
\end{figure*}

The experiments are repeated five times with different random seeds for each combination of the adaptation method, point selection strategy, and target area. The average adaptation performance when methods use only 400 labeled pixels -- close to the number labeled by users in online studies -- is shown in Table \ref{table:active-learning-table}, while the variation in accuracy across the whole range of additional training points is shown in Figure \ref{fig:offline-results}. (See also Appendix F for the full set of curves.)
\begin{table}[]
\caption{Results of fine-tuning on 400 points selected by different query methods, averaged over four target areas and five random seeds.}
\label{table:active-learning-table}
\centering
\resizebox{\textwidth}{!}{%
\begin{tabular}{@{}l|rr|rr|rr|rr|rr@{}}
\toprule
& \multicolumn{2}{c|}{\textbf{Last 1 Layer}} & \multicolumn{2}{c|}{\textbf{Last 2 Layers}} & \multicolumn{2}{c|}{\textbf{Last 3 Layers}} & \multicolumn{2}{c|}{\textbf{Group Params}} & \multicolumn{2}{c}{\textbf{Dropout}} \\
Query method & \multicolumn{1}{c}{Acc} & \multicolumn{1}{c|}{IoU} & \multicolumn{1}{c}{Acc} & \multicolumn{1}{c|}{IoU} & \multicolumn{1}{c}{Acc} & \multicolumn{1}{c|}{IoU} & \multicolumn{1}{c}{Acc} & \multicolumn{1}{c|}{IoU} & \multicolumn{1}{c}{Acc} & \multicolumn{1}{c}{IoU} \\ \midrule
\textbf{Baseline} &  0.725 &  0.510 &  0.725 &  0.510 &  0.725 &  0.510 &  0.725 &  0.510 &  0.725 &  0.510\\
\hline 
\textbf{Random} & 0.806 & 0.608 & 0.825 & 0.677 & 0.824 & 0.658 & 0.791 & 0.562 & 0.787 & 0.597  \\
\textbf{Entropy} & 0.736 & 0.501 & 0.731 & 0.587 & 0.765 & 0.572 & 0.760 & 0.520 & 0.741 & 0.550  \\
\textbf{Min-Margin} & 0.811 & 0.608 & 0.834 & 0.701 & 0.832 & 0.685 & 0.793 & 0.580 & 0.785 & 0.601  \\ \midrule
\textbf{Mistakes} & 0.729 & 0.551 & 0.781 & 0.631 & 0.756 & 0.621 & 0.787 & 0.575 & 0.762 & 0.609\\
 \bottomrule
\end{tabular}%
}
\end{table}

As the \textbf{Last 1 Layer} and \textbf{Last 2 Layers} fine-tuning methods tend to perform best, we chose them for use in online experiments.

For all fine-tuning methods, we observed a similar ranking of the performance of active learning query methods, with \textbf{Min-Margin} performing best, but only slightly better than \textbf{Random}, and \textbf{Entropy} performing worst. Most interestingly, the \textbf{Mistakes} method performs significantly worse than \textbf{Random}: even giving the model access to ground truth knowledge does not improve performance. On the other hand, in online experiments (Sec.~\ref{sec:experiments_online}), we show that replacing this mock ``uniform teacher'' with a human teacher \emph{does} improve performance.


\section{Online study of the hybrid labeling system} \label{sec:experiments_online}

As can be seen from the indicated confidence intervals in Fig. \ref{fig:offline-results}, it is not clear that we can expect any of the active learning methods to outperform random selection of data points to label, as was previously often observed in active learning literature \cite{settles2011theories}. Thus, we test our hybrid labeling system -- the \textbf{Human} query method -- against the \textbf{Random} point query method.

As explained in Sec.~\ref{sec:study_design}, we use a standard crowdsourcing setup in the Amazon Mechanical Turk system to acquire unbiased ground truth labels in four different 85km$^2$ \textit{target areas} in New York. We collected a total of 6009 labels on randomly selected points, from 54 unique labelers, resulting in a dataset of 3441 unambiguous labeled points. These labels agree with the Chesapeake ground truth data \cite{chesapeakeData1} $91.1\%$ of the time, which is in line with that data product's published quality estimates \cite{chesapeakeData2}.

Second, we use these high-quality labels to evaluate our hybrid system as described in Section \ref{sec:study_design}. We had 50 workers\footnote{See Appendix B for study details.} utilize our web tool to fine-tune the pretrained baseline model in a series of 15 minute sessions on four target areas using two adaptation methods.
In each session, every time the user induces retraining of the model, we calculate that model's performance on the set of crowdsourced ground truth labels from the area in which they are working. We compare this method with the \textbf{Random} query method using the crowdsourced ground truth dataset. In the crowdsourced labeling task, users take $\sim 3$ seconds to label each pixel they are shown. Thus, in a 15-minute window, they could provide labels on $\sim 300$ randomly sampled points. A central question is that of \textit{label efficiency}: is human time and money best spent by labeling the central pixels of random patches of aerial imagery (human as \emph{label oracle}) or by using our interactive tool (human as \emph{query method} and \emph{label oracle})?

\begin{figure*}[t]
    \centering
    \includegraphics[height=3.5cm]{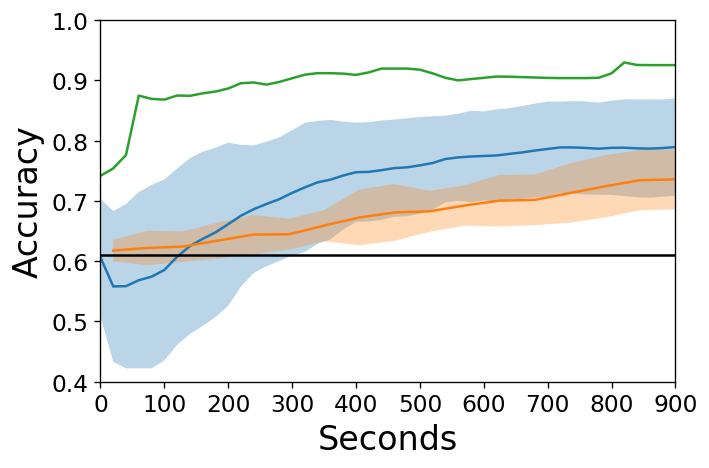}
    \includegraphics[height=3.5cm]{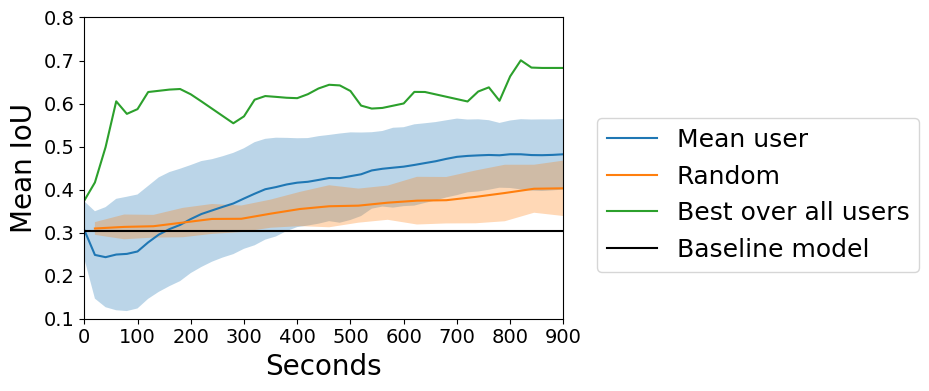}
    \caption{Performance of \textbf{Human} and \textbf{Random} query methods for model fine-tuning in a 15 minute time window, measured in pixel accuracy (left) and mean IoU (right). Mean user performance is calculated over the top 50\% of users and considers sessions using the \textbf{Last 2 layer} fine-tuning method. Random performance is averaged over 10 seeds, with points assumed to be added every 3 seconds. Both methods are averaged over the same four \textit{target areas}.}
    \label{fig:online-results}
\end{figure*}

The subplots in Figure \ref{fig:online-results} show accuracy and mean intersection-over-union of intermediate models achieved at different times in the 15 minute fine-tuning sessions, averaged across users. In the case of the \textbf{Random} method, we assume that 300 points are added at uniform time intervals and the model is retrained every 45 seconds. As the model for a specific user will fluctuate in performance over the duration of a single session -- users pick up on different deficiencies in the core model at different points during a session -- we summarize the \textbf{Human} method as a whole by averaging performance metrics over sets of users. 
Models fine-tuned using the \textbf{Human} query method consistently outperform models that are fine-tuned with \textbf{Random} queried points, within 3 minutes of labeling ($\sim60$ samples). 
The top curve in Fig.~\ref{fig:online-results} shows the best model over all users at each point in time. We find that users' area-adjusted performance in task 2 is highly predictive of their performance in task 3 ($p<0.01$, rank-correlation $\rho=0.4$): of the top 25 (half) of users ranked by (IoU) performance in task 2, 17 are also among the top 25 in task 3. Thus, \textbf{the better-performing labelers are detectable} in a statistically significant manner. This indicates that the users are developing different levels of intuition about the inner workings of the network and the fine-tuning method. In addition, the performance of the users in tasks 2 and 3 is far less predictive ($\rho=0.1$) of their performance in task 4, where the fine-tuning method is switched. This indicates that users are building a \textbf{theory of mind for the AI agent} they work with in tasks 1-3, which is then broken as by a slight change in the learning algorithm in task 4.

Our offline experiments with the \textbf{Mistake} method indicate that the model simply knowing where its errors are cannot automatically beat the \textbf{Random} selection of points for labeling. This indicates that \textbf{human guidance goes beyond simply quickly spotting errors}, especially for best performers, reminiscent of the super-teacher idea \cite{superteacher}. Text feedback from users (see Appendix C) provides further interesting insights that should be useful in the design of hybrid systems of this kind.

\section{Conclusions} \label{sec:discussion}

We have conducted a study of hybrid human-AI intelligence on the task of high-resolution land cover mapping. We demonstrate that giving control of the data selection process to the human yields significant improvements in model accuracy. Our user studies show that users develop a theory of mind for the ML system, learning to understand the workings of particular AI algorithms with variation in this skill correlated across different tasks. If we consider this learning framework in the context of usual ML challenges, where humans (engineers/researchers) are paired with machines (new model designs), we can see that similar variation in accuracy comes simply from humans' knack for the particular challenge, even when everyone uses the same architecture and learning algorithms.

By injecting the human into the learning loop, gains from both the human and the AI labor are amplified, not replaced. For the machine, sparse but well-chosen human feedback reduces the cost of computational resources needed to adapt models. For the human, increased sample efficiency of the ML systems acts like an ever-more useful wand with which they can paint the land cover. Together, this collaboration achieves critical cost reduction in practical problems. The Chesapeake dataset was created in 10 months at a cost of \$1.3 million, though it covers just 2\% of the US~\cite{chesapeakeData2} with an estimated accuracy of 90\%-95\%. The best user from our study, averaged over the 4 target areas, achieved an accuracy of 89.1\% in just one hour of labeling work.\footnote{This number is in line with the recent state-of-the-art algorithm\cite{malkin2018label} which uses 30m low-resolution labels as additional data. Our approach does not rely on the existence of such low-res labels.} If such users were to label the entire Chesapeake Bay watershed using our method, this would take 925 hours of work at a cost of \$18.5k. Of course, other tradeoffs between accuracy and cost are possible by allowing users to work longer on each area or even to work collaboratively.

In problems of massive scale where unlabeled data is practically limitless, such as land cover labeling, it is not likely that a few months of labeling through our tool would create enough training data that the need for human labor would disappear. Instead, applications that are now infeasible, such as quick generalization to new areas or addition of new target classes (shown in Appendix D), would become feasible, making both the ML algorithms and human labor more valuable than before. 

{
\small
\bibliographystyle{plain}
\bibliography{citations}
}

\appendix

\clearpage
\newpage

Supplementary video files can be downloaded here: \url{https://www.dropbox.com/s/nh4wlhf6ez2a17a/neurips_2019_si.zip?dl=0}.

\section{Model and training details}

Our core U-net model is an encoder-decoder network architecture with skip connections. It contains four down-sampling and four up-sampling layers. For down-sampling, we use a simple 2$\times$2 max-pooling. For up-sampling, we use deconvolution (transposed convolution). Before each down-sampling and up-sampling layer, we insert two convolutional layers. The first two convolutional layers have 32 $3\times3$ filters. Group normalization \cite{wu2018group} is applied after the second convolution in every layer followed by ReLU. Valid padding is used in all layers making the predicted output smaller than the input. The number of filters is doubled after each pooling layer, the representational bottleneck layers use 512 $3\times 3$ filters. We trained the network for 100 epochs $\sim90000$ randomly selected image patches of size 240 x 240 sampled from the state of Maryland. We used the Adam optimizer \cite{kingma2014adam} with cross-entropy as segmentation loss and an initial learning rate of $0.001$ decaying to $0.0001$ after 60 epochs. 

The \textbf{Last $k$ Layers} and \textbf{Group Params} methods were all trained using the Adam optimizer for 10 epochs, $\epsilon = 10^{-5}$. Learning rates were set as follows: 0.01 for last $1$ layer, 0.005 for last $2$ layers, 0.001 for last $3$ layers, and $0.0025$ for group parameters.

\section{Online study setup}
We deployed both our tool for collecting unbiased randomly sampled land cover data (see Section \ref{sec:random_tool}), and an online user study (described in the main task) on Amazon Mechanical Turk. Our participants were compensated at a fair wage of \$20 per hour and were screened according to the following qualifications: hold above a 95\% approval rating on the platform, unique study participant, and based in the United States. While the hourly rate for these workers was \$20, each worker used a set of unique web links that communicated to a dedicated server running 4 GPUs. The cost of computation was thus negligible compared to the cost of labor. This was deliberate, as the main reason for the development of the tool was to allow program managers in various agencies to explore by themselves the potential of land cover mapping in their applications \emph{before} committing large human or computational resources to it. The computational throttle, however, does influence the design: The users have to test a small 500 $\times$ 500 patch of land at a time in search of areas to correct, and retraining has to be explicitly initiated but the user, rather than running continuously in the background (See the attached tutorial video used in the study, for which updated codecs may be needed). Optimal distribution of computational and human labor costs will always depend on a variety of social and practical/engineering factors. We plan on investigating other tradeoffs and systems, esp. in collaborative settings where multiple humans work on the same map, with artificially intelligent amplification that uses larger computational resources.  

Although the tool had a variety of functionalities to speed up the task, such as control of transparency, and training initiation with hot keys, we consciously tried to reduce the complexity so that the users would be proficient with the tool after the first 15-minute task. Thus we omitted some functions which would be useful and necessary in larger deployments, such as the various type of ``undo'' and ``revert'' functionalities allowing users to backtrack after, 'for example, accidentally providing incorrectly labeled pixels, as these can be inferred after the study by an observed drop in accuracy after a large disagreement in labels. 
Similarly, for simplicity, and to separate the individual ability to use complex labeling tools to label more or less of the land from the user's ability to reason over what to label, the users were only required to make point labels as the means of communicating with the tool, as opposed to some other popular approaches to crowd-sourced segmentation where the users can use brushes or draw polygons. As can be seen in the video, a couple of clicks followed by retrain are sufficient to correct large pieces of land, any how, and more detailed segmentation tools could sidetrack some users in the series of short tasks.

In the online user study, each task contained the same layout: every participant was presented with a set of written instructions explaining how the web interface worked. A video tutorial (see download link above Appendix A) was also provided to demonstrate the labeling task in an separate example area. Participants were asked to spend 15 minutes each correcting what they perceived to be incorrect labels on 4 areas (1 at a time), with server validation ensuring that proper time and non-duplicate results were recorded. Afterwards, participants were asked to provide feedback to verbal questions asking whether they believed their own performance improved, whether they believed the model performance improved, and whether their personal labeling strategy evolved in the course of completing the different areas. Specifically, users were asked ``How well did you feel the AI model did in responding to labels provided by you?'', ``Did you change your labeling strategy as you spent more time using the tool? If so, how?'', and for further comments/feedback. See Section \ref{sec:feedback}. These questions were a way for us to gauge the level of user's acceptance of the tool's classification, in addition to the actual accuracy, given that predictions from an AI black box are often mistrusted. Our hypothesis was that our tool would make the ML algorithm less of a black box to the users and more trustworthy.

\subsection{Random labeling tool} \label{sec:random_tool}
\begin{figure}[h]
\centering
\includegraphics[width=\textwidth]{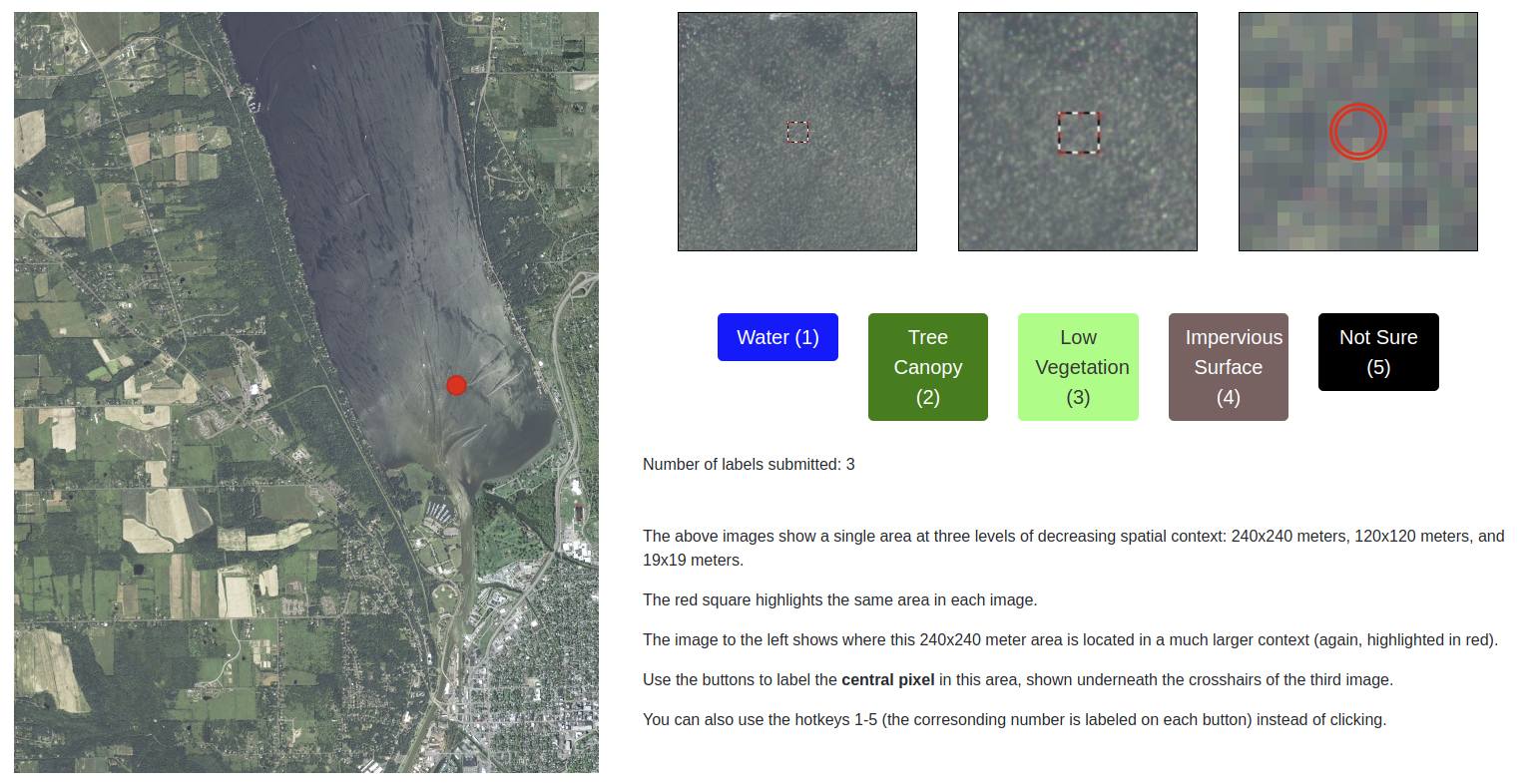}
\caption{Interface of our random land cover labeling tool.}
\label{fig:random_tool}
\end{figure}

We designed a tool for collected unbiased labels from users, shown in Figure \ref{fig:random_tool}. The user is shown a patch of imagery from one of the target areas and asked to label the central point. The patch is shown simultaneously at 3 levels of zoom. The user is allowed to avoid labeling ambiguous points, such as the ones that fall on the boundary between two land cover classes.

When multiple users disagree on the label of a point, or where the user's label disagrees with the baseline ground truth \cite{chesapeakeData1,chesapeakeData2} used in offline experiments, the point is sent back into the labeling queue to be evaluated by 5 more labelers and finally labeled by the majority vote.

\section{Human-AI Relationship}
\label{sec:feedback}

Qualitative feedback from the users in the online study suggests other interesting features of the results, particularly as it relates to their conceptualization of a optimal working relationship with the AI model:
\begin{itemize}
    \item Average user performance will \textit{degrade} below the baseline performance of the core model in the first few minutes while users initially attempt to fine-tune the model to work in the first geographic area they were attracted to. According to one user, ``\textit{I would spend a lot of time in one area trying to detail everything possible}''.
    \item Text feedback suggests that the model's fine-tuning behavior was indeed learned by some users, for example: \textit{``As I got more experienced with the tool I would map broad types of the environment at once, for example tree canopies, building structure, water, etc\dots''} and \textit{``towards the end I was more or less trying to correct certain areas by giving the AI a smaller number of samples from the spot that was incorrect and other spots that it should be labeled like''}. In other cases, this behavior was a by-product of human biases: \textit{``I wanted a little bit of all types in the outline so I chose a place with bridge, water, trees and vegtations as it was more interesting for me than just all of one type''} [sic]. 
    
    Users are biased towards labeling pixels with certain classes over others. We note that in the first \textit{target area}, the ground truth label distribution is: 8.5\% water, 53.9\% tree canopy,	35.6\% low vegetation, and 2.0\% impervious surfaces,  while the average distribution of labels that the users submitted is much more balanced: 18.8\% water, 29.8\% tree canopy, 23.2\% low vegetation, and 28.2\% impervious.
    \item Users developed strategies and mental frameworks for how they related to the tool, ranging from changes in their own confidence in successfully providing good labels to the model to notions of responsibility in completing a hybrid task \textit{``The interface responded well to labels \dots unfortunately it also responded very well to mistakes on my part''}, \textit{``On the first [task], I was honestly lost and scared I was screwing up''}.
    \item Users enjoyed using the tool, suggesting that interacting with a model that is adapting to user feedback is a valuable experience for the human: \textit{``It was mostly accurate. Almost magical''}, \textit{``This was fun and engaging''}, and \textit{``This was a great way to quickly get a very refined map. The video was extremely helpful!''}.
    
    \item Users had more confidence in the ability of the AI model than what was communicated to them (which was very little). When users failed to see substantive improvements in the model performance, they tended to blame themselves rather than the model for the poor performance: \textit{``In the beginning I felt I was doing something wrong because it didn't seem to be responding correctly.''}
\end{itemize}

\section{Online class additions}
\begin{figure}[ht]
\centering
\includegraphics[width=0.5\textwidth]{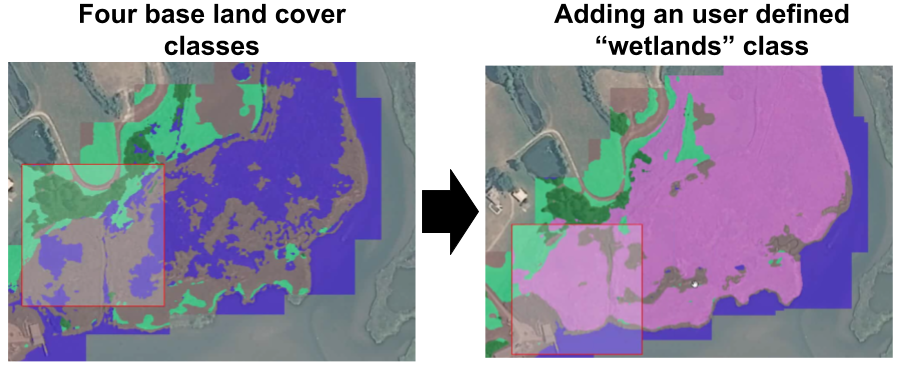}
\caption{Screenshots from the web-tool that show how users can define new classes through adding point examples. The area shown in the Figure is not well described by any of the land cover classes that the base model is trained on and can be split off into a ``wetlands'' class entirely through using the web interface.}
\label{fig:wetland_tool}
\end{figure}

One important feature of our fine-tuning interface is that it can enable users to define \textit{new} land cover classes on-the-fly by providing point examples from these classes. Fig. \ref{fig:wetland_tool} shows a before and after example of this process: a ``wetlands'' area is originally labeled as a mix of the \textit{built}, \textit{water}, and \textit{field} classes, however after an user defines a new \textit{wetlands} class by adding appropriate samples from this area, the model is able to generalize over the entire feature. This functionality is also demoed in a video packaged in this Appendix (see download link above Appendix A).

\section{User attention}
\begin{figure}[ht]
\centering
\includegraphics[width=1\textwidth]{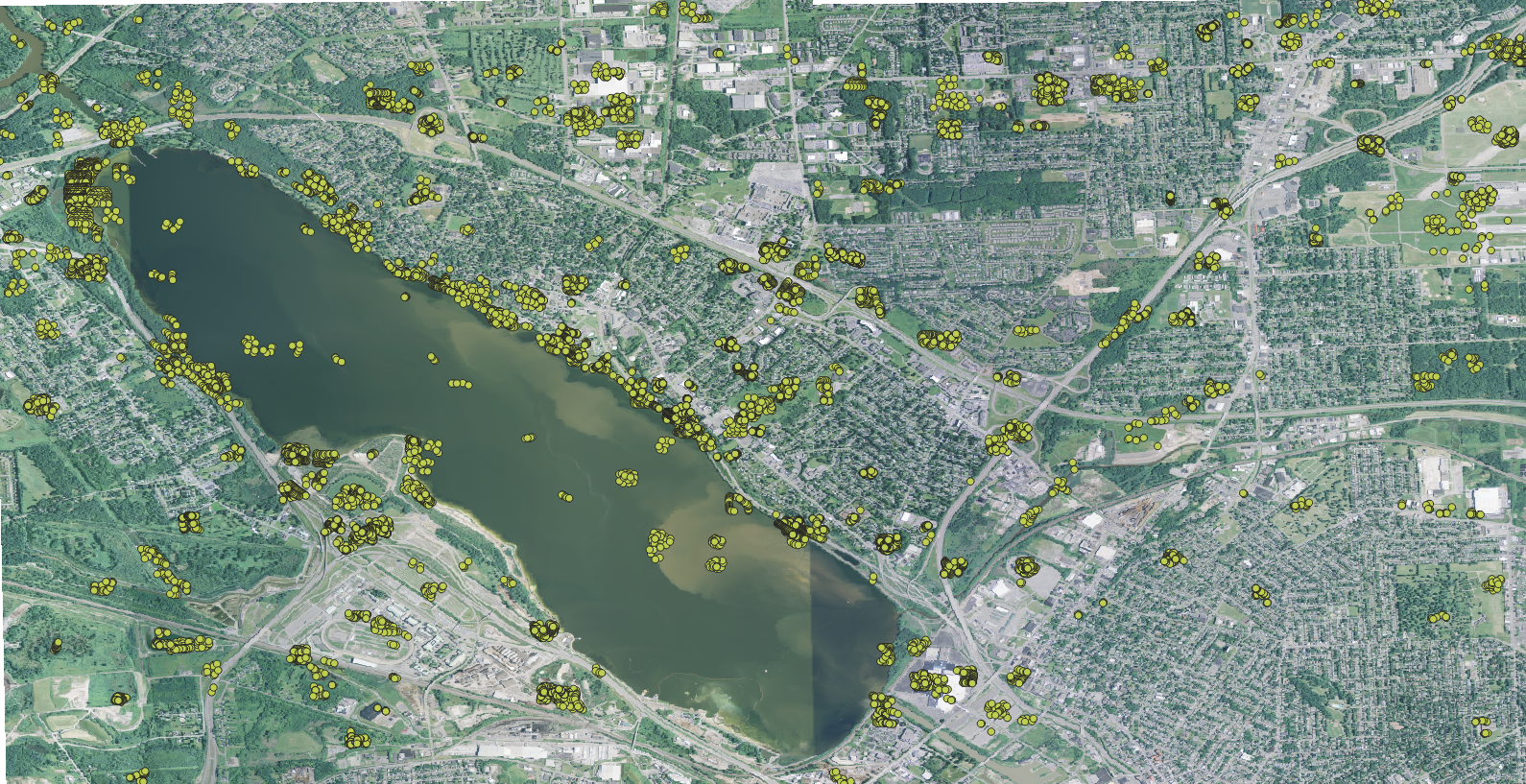}
\includegraphics[width=1\textwidth]{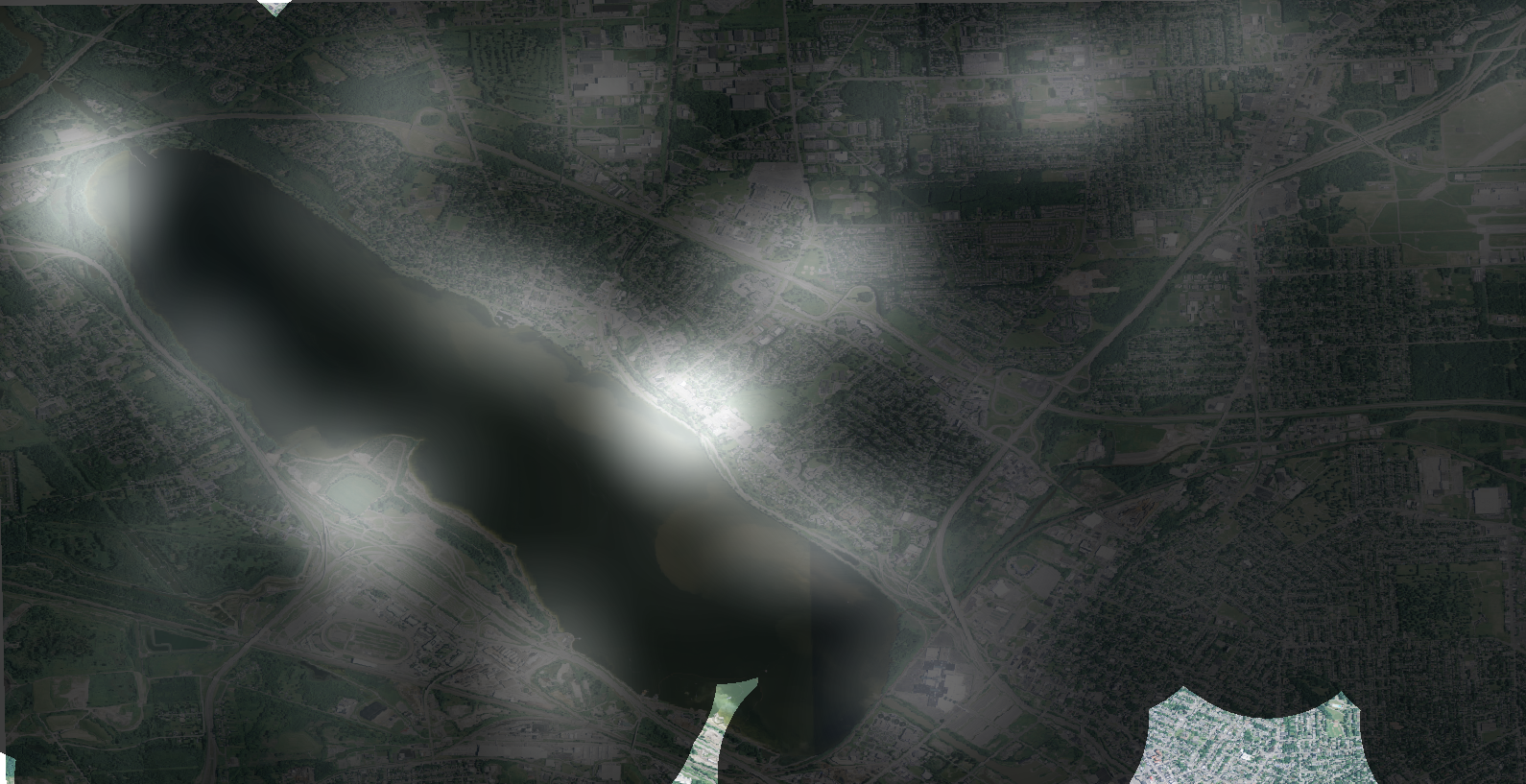}
\caption{(\textbf{Top}) Visualization of the locations of all user-submitted labels through the web-tool in the fourth testing area. (\textbf{Bottom}) Kernel density estimation surface fit using the above points. Users focus on fine-tuning the model in similar locations, while completely ignoring other locations. This highlights the data biases that users incorporate into the fine-tuning process.}
\label{fig:heatmap}
\end{figure}

We gather the locations where all users submitted label corrections when using the web-tool, and visualize these for one of the study areas in Figure \ref{fig:heatmap}. Here, we can see that users behavior is distinctively not random - there are common locations in the study area that users are drawn to, while there are other locations that are ignored by all users. We visualize the density of these user submitted points with a kernel density estimation (KDE) surface and find that the shoreline of the prominent lake in the study area contains the most popular locations where users submit corrections. On the other hand, there is a standard residential neighborhood in the south east corner of the area that is not visited by any user. Qualitatively, users are drawn to visually striking features and ignore more mundane features.

\section{Full performance graphs}

\captionsetup[figure]{font=small,skip=0pt}
\captionsetup[subfigure]{font=small,skip=0pt}
\begin{figure*}[htpb]
        \centering
        
        	\begin{subfigure}[b]{.27\textwidth}
        		\includegraphics[width=\textwidth]{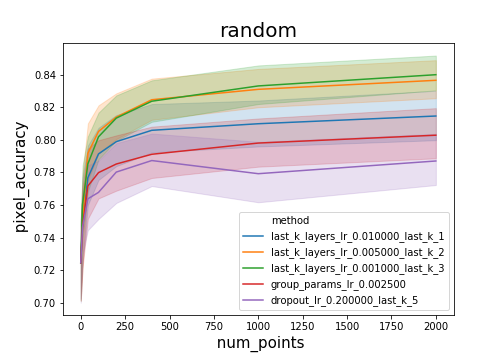}
        		\caption{Random Pixel Acc.}
        		\label{fig:sub1}
        	\end{subfigure}%
        	\begin{subfigure}[b]{.27\textwidth}
        		\includegraphics[width=\textwidth]{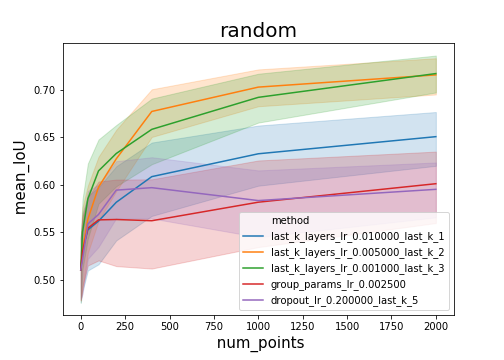}
        		\caption{Random Mean IoU.}
        		\label{fig:sub2}
        	\end{subfigure}
        	\begin{subfigure}[b]{.27\textwidth}
        		\includegraphics[width=\textwidth]{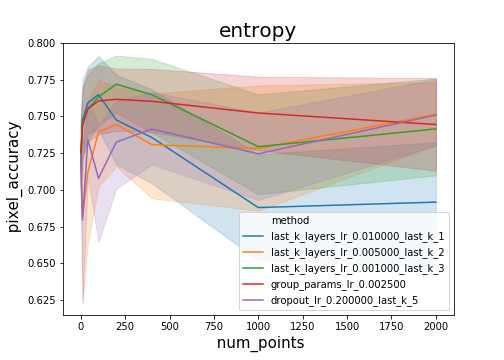}
        		\caption{Entropy Pixel Acc.}
        		\label{fig:sub3}
        	\end{subfigure}
        	\begin{subfigure}[b]{.27\textwidth}
        		\includegraphics[width=\textwidth]{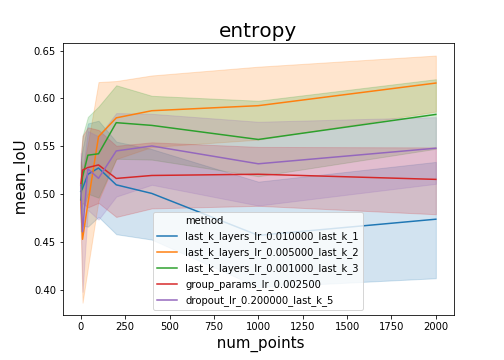}
        		\caption{Entropy Mean IoU.}
        		\label{fig:sub4}
        	\end{subfigure}
        	\begin{subfigure}[b]{.27\textwidth}
        		\includegraphics[width=\textwidth]{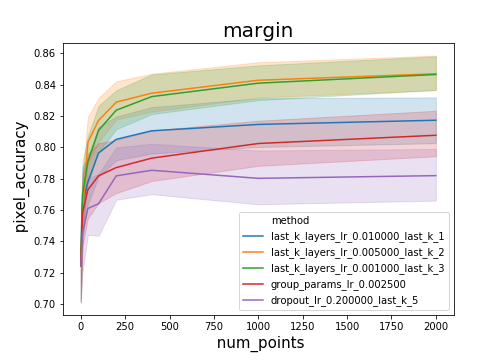}
        		\caption{Min Margin Pixel Acc.}
        		\label{fig:sub5}
        	\end{subfigure}
        	\begin{subfigure}[b]{.27\textwidth}
        		\includegraphics[width=\textwidth]{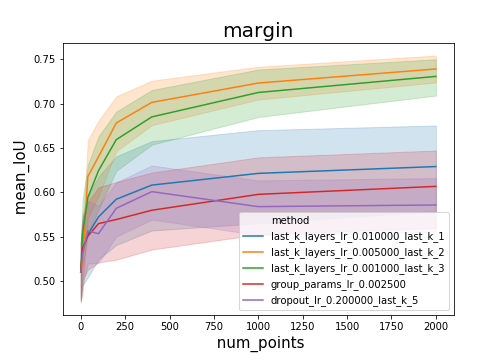}
        		\caption{Min Margin Mean IoU.}
        		\label{fig:sub6}
        	\end{subfigure}
    	\caption{ Full Performance Data Query Methods }
    	\label{fig:data-query}
    \end{figure*}

\clearpage
\newpage

\addtolength{\textheight}{3in}

\begin{figure*}[!tbp]
        \centering
        \captionsetup[figure]{font=small,skip=0pt}
        \captionsetup[subfigure]{font=small,skip=0pt}
        	\begin{subfigure}[b]{.36\textwidth}
        		\includegraphics[width=\textwidth]{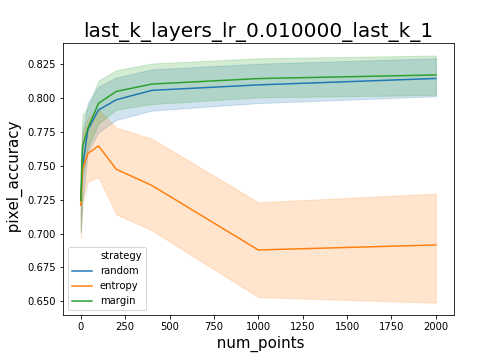}
        		\caption{Last Layer Pixel Acc.}
        		\label{fig:sub1}
        	\end{subfigure}%
        	\begin{subfigure}[b]{.36\textwidth}
        		\includegraphics[width=\textwidth]{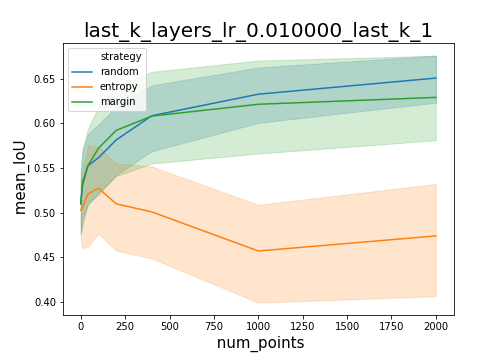}
        		\caption{Last Layer Mean IoU}
        		\label{fig:sub2}
        	\end{subfigure}
        	\begin{subfigure}[b]{.36\textwidth}
        		\includegraphics[width=\textwidth]{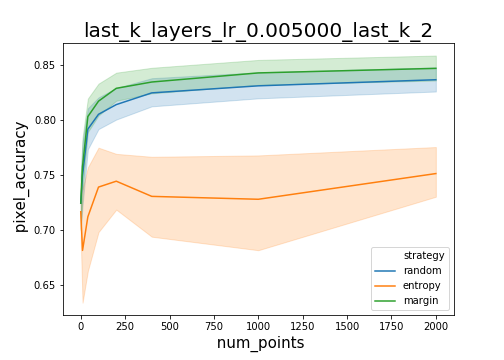}
        		\caption{Last 2 Layers Pixel Acc.}
        		\label{fig:sub3}
        	\end{subfigure}
        	\begin{subfigure}[b]{.36\textwidth}
        		\includegraphics[width=\textwidth]{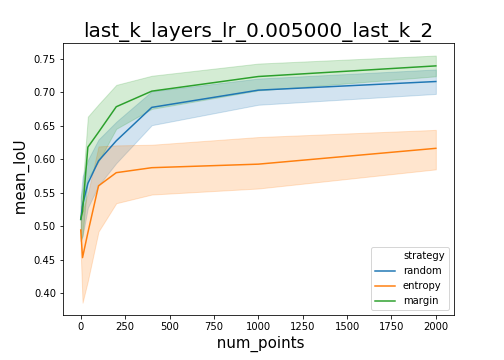}
        		\caption{Last 2 Layers Mean IoU}
        		\label{fig:sub4}
        	\end{subfigure}
        	\begin{subfigure}[b]{.36\textwidth}
        		\includegraphics[width=\textwidth]{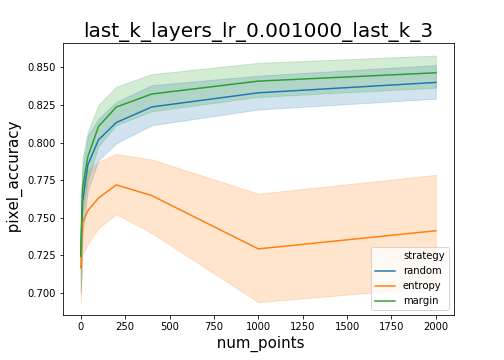}
        		\caption{Last 3 Layers Pixel Acc.}
        		\label{fig:sub5}
        	\end{subfigure}
        	\begin{subfigure}[b]{.36\textwidth}
        		\includegraphics[width=\textwidth]{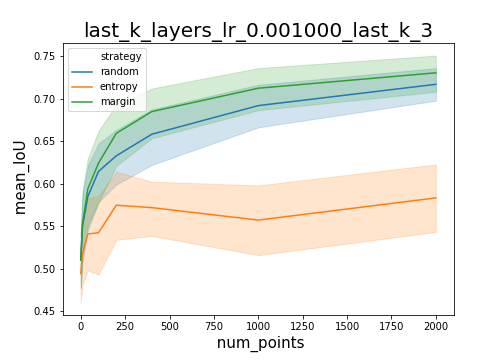}
        		\caption{Last 3 Layers Mean IoU}
        		\label{fig:sub6}
        	\end{subfigure}
        	
        	\begin{subfigure}[b]{.36\textwidth}
        		\includegraphics[width=\textwidth]{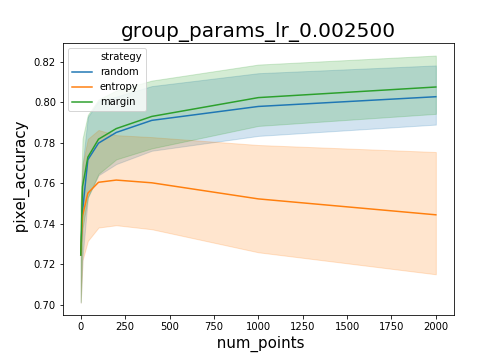}
        		\caption{Group Params Pixel Acc.}
        		\label{fig:sub7}
        	\end{subfigure}
        	\begin{subfigure}[b]{.36\textwidth}
        		\includegraphics[width=\textwidth]{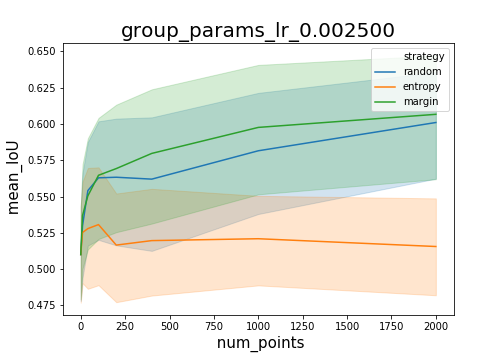}
        		\caption{Group Params Mean IoU}
        		\label{fig:sub8}
        	\end{subfigure}
        	
        	\begin{subfigure}[b]{.36\textwidth}
        		\includegraphics[width=\textwidth]{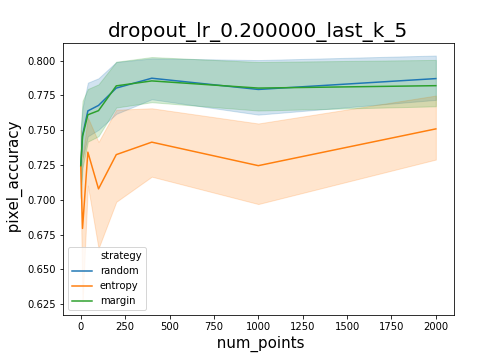}
        		\caption{Dropout Pixel Acc.}
        		\label{fig:sub9}
        	\end{subfigure}
        	\begin{subfigure}[b]{.36\textwidth}
        		\includegraphics[width=\textwidth]{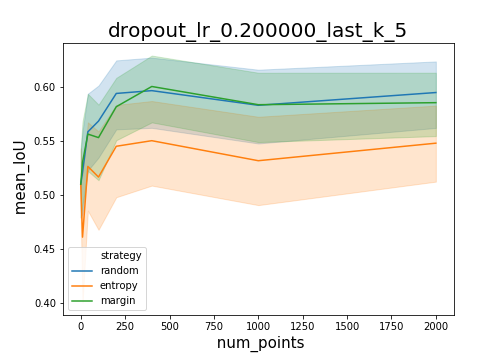}
        		\caption{Dropout Mean IOU}
        		\label{fig:sub10}
        	\end{subfigure}
    	\caption{ Full Performance Fine-tuning Methods. }
    	\label{fig:finetuning}
    \end{figure*}

\end{document}